\documentclass[epj,twocolumns]{svjour_mod}
\usepackage[utf8]{inputenc}
\usepackage{latexsym}
\usepackage{graphics}
\usepackage{amssymb}
\usepackage[colorlinks,citecolor=blue,urlcolor=blue,linkcolor=blue]{hyperref}
\usepackage{hyphenat}
\usepackage{amsmath}
\usepackage{cite}
\usepackage{relsize}
\usepackage{trimclip}
\usepackage{booktabs} 
\usepackage{tabularx}
\usepackage{multirow}
\usepackage{fixmath}
\usepackage{tikz}
\usetikzlibrary{calc}
\usepackage{placeins}
\usepackage{xspace}
\usepackage{slashed}

\DeclareRobustCommand{\ensuremathrm}[1]{\ensuremath{\mathrm{#1}}\xspace}

\DeclareRobustCommand{\rd}{\ensuremathrm{d}} 

\DeclareRobustCommand{\GeV}{\ensuremathrm{GeV}\xspace}
\DeclareRobustCommand{\TeV}{\ensuremathrm{TeV}\xspace}

\DeclareRobustCommand{\nnlojet}{\mbox{\textsc{\tt NNLOjet}}\xspace}
\DeclareRobustCommand{\radish}{\mbox{\textsc{\tt RadISH}}\xspace}
\DeclareRobustCommand{\jet}{\text{jet}\xspace}

\DeclareRobustCommand{\mll}{\ensuremath{M_{\ell\ell}}\xspace}
\DeclareRobustCommand{\mlnu}{\ensuremath{M_{\ell\nu}}\xspace}

\newcommand{\as}{\alpha_\mathrm{s}}

\newcommand{\NNNLL}{\text{N${}^3$LL}}

\newcommand{\pt}{p_\perp}

\begin{document}

\title{The transverse momentum spectrum of weak gauge bosons at \NNNLL+NNLO}

\author{Wojciech Bizo\'{n}\inst{1,2}
\and Aude Gehrmann-De Ridder\inst{3,4}
\and Thomas Gehrmann\inst{4}
\and Nigel Glover\inst{5}
\and Alexander Huss\inst{6}
\and Pier Francesco Monni\inst{6}
\and Emanuele Re\inst{6,7}
\and Luca Rottoli\inst{8,9}
\and Duncan M. Walker\inst{5}}

\institute{
  Institut f{\"u}r Theoretische Teilchenphysik (TTP), KIT, 76128 Karlsruhe, Germany
  \and
  Institut f{\"u}r Kernphysik (IKP), KIT, 76344 Eggenstein-Leopoldshafen, Germany
\and Institute for Theoretical Physics, ETH, CH-8093 Z\"urich,
  Switzerland
\and Department of Physics, University of Z\"urich, CH-8057 Z\"urich, Switzerland
\and Institute for Particle Physics Phenomenology, Durham University,
Durham DH1 3LE, UK 
\and CERN, Theoretical Physics Department, CH-1211 Geneva 23, Switzerland 
\and LAPTh, Universit\'e Grenoble Alpes, Universit\'e Savoie Mont Blanc, CNRS, 74940 Annecy, France
\and Dipartimento di Fisica G. Occhialini, U2, Universit\`a degli Studi di Milano-Bicocca,
Piazza della Scienza, 3, 20126 Milano, Italy
\and INFN, Sezione di Milano-Bicocca, 20126, Milano, Italy}

\date{}

\abstract{We present accurate QCD predictions for the transverse
  momentum ($\pt$) spectrum of electroweak gauge bosons at the LHC for
  $13~\TeV$ collisions, based on a consistent combination of a NNLO
  calculation at large $\pt$ and N$^3$LL resummation in the small
  $\pt$ limit. The inclusion of higher order corrections leads to
  substantial changes in the shape of the differential distributions,
  and the residual perturbative uncertainties are reduced to the few
  percent level across the whole transverse momentum spectrum.
  We examine the ratio of $\pt$ distributions in charged- and
  neutral-current Drell-Yan production, and study different
  prescriptions for the estimate of perturbative uncertainties that
  rely on different degrees of correlation between these processes. We
  observe an excellent stability of the ratios with respect to the
  perturbative order, indicating a strong correlation between the
  corresponding QCD corrections.}


\maketitle

\section{Introduction}
\label{sec:intro}
The differential spectrum of electroweak gauge bosons, measured via
their leptonic decays, is among the most prominent observables at the
LHC.

Owing to the outstanding precision of their experimental
measurement~\cite{Chatrchyan:2011wt,Aad:2012wfa,Aad:2014xaa,Aad:2015auj,Aaij:2015gna,Khachatryan:2015oaa,Khachatryan:2015paa,Aaij:2015zlq,Aad:2016izn,Khachatryan:2016nbe,Aaij:2016mgv,Sirunyan:2017igm,Aaboud:2017svj,Aaboud:2017ffb},
such observables allow for a precise extraction of some of the
Standard Model (SM) parameters --- such as the $W$ boson
mass~\cite{Aaboud:2017svj}, or parton
densities~\cite{Ball:2017nwa,Boughezal:2017nla,Bacchetta:2017gcc,Bertone:2019nxa}
--- as well as for the calibration of widely used event generators and
analysis tools.
For this reason, an accurate theoretical understanding of such
observables is paramount to exploit the precise data and perform
meticulous tests of the SM.

Inclusive and differential distributions for neutral and charged
Drell-Yan (DY) production with lepton pair invariant mass $M$ are
nowadays known with very high precision. The total cross section is
known fully differentially in the Born phase space up to
NNLO~\cite{Hamberg:1990np,vanNeerven:1991gh,Anastasiou:2003yy,Melnikov:2006di,Melnikov:2006kv,Catani:2010en,Catani:2009sm,Gavin:2010az,Anastasiou:2003ds},
while differential distributions in transverse momentum $\pt$ were
recently computed up to NNLO both for
$Z$-~\cite{Ridder:2015dxa,Ridder:2016nkl,Gehrmann-DeRidder:2016jns,Gauld:2017tww,Boughezal:2015ded,Boughezal:2016isb}
and
$W$-boson~\cite{Boughezal:2015dva,Boughezal:2016dtm,Gehrmann-DeRidder:2017mvr}
production.  In the DY distributions, electroweak corrections become
important especially at large transverse momenta, and they have been
computed to NLO accuracy
in~\cite{Kuhn:2005az,Kuhn:2007qc,Denner:2009gj,Denner:2011vu}.

In kinematical regimes dominated by soft and collinear radiation, the
fixed-order perturbative series for the differential $\pt$
distribution is affected by large logarithmic terms of the form
$\as^n L^{2n-1}/\pt$, with $L\equiv\ln(M/\pt)$, which must be resummed
to all orders for a reliable theoretical prediction.
In such regimes, the perturbative (logarithmic) accuracy is defined in
terms of the \emph{logarithm} of the cumulative cross section ${\rm \Sigma}$
as
\begin{align}
\label{eq:cumulant-initial}
  \ln &\left({\rm \Sigma}(\pt)\right) \equiv \ln\left(\int_0^{\pt} \rd
        \pt' \; \frac{\rd {\rm \Sigma}(\pt')}{\rd \pt'} \right)\notag\\
&= \sum_n \left\{{\cal O}\left(\as^nL^{n+1}\right) + {\cal
  O}\left(\as^nL^{n}\right) + \dots\right\}.
\end{align}
One refers to the dominant terms $\as^n L^{n+1}$ as leading
logarithmic (LL), to terms $\as^n L^{n}$ as next-to-leading
logarithmic (NLL), to $\as^n L^{n-1}$ as next-to-next-to-leading
logarithmic (NNLL), and so on.
The resummation of the $\pt$ spectrum of SM bosons has been studied in
a multitude of theoretical formulations throughout the
years~\cite{Parisi:1979se,Collins:1984kg,Balazs:1997xd,Landry:2002ix,Becher:2010tm,Bozzi:2010xn,Becher:2011xn,GarciaEchevarria:2011rb,Monni:2016ktx,Ebert:2016gcn,Bizon:2017rah},
and the current state of the art for phenomenological studies at the
LHC reaches N$^3$LL
accuracy~\cite{Bizon:2017rah,Chen:2018pzu,Bizon:2018foh,Becher:2019bnm}.

In this article, we reach a new milestone in the theoretical 
description of transverse momentum distributions in both neutral
and charged DY production, aiming for per-cent level precision
throughout the full kinematical range. This is achieved by matching
the fixed-order NNLO QCD predictions with the \NNNLL{} resummation of
large logarithmic corrections.  We adopt the momentum-space
formulation of refs.~\cite{Monni:2016ktx,Bizon:2017rah}, in which the
resummation is performed by generating the QCD radiation by means of a
Monte Carlo (MC) algorithm. %
All the necessary ingredients for the \NNNLL{} $\pt$ resummation have
been computed in
refs.~\cite{Catani:2012qa,Gehrmann:2014yya,Echevarria:2016scs,Li:2016ctv,Vladimirov:2016dll,Moch:2018wjh,Lee:2019zop}.
The combined framework enables fully differential \NNNLL+NNLO
predictions for distributions that take proper account of the fiducial
volume definitions used in the experimental measurements.

The article is organised as follows. In section~\ref{sec:calculation}
we briefly review the computation of the NNLO differential
distributions in DY-pair production with the parton-level code
\nnlojet, as well as the resummation for the $\pt$ distributions using
the computer program \radish. Section~\ref{sec:results} describes our
results for $13~\TeV$ LHC collisions. Finally,
section~\ref{sec:conclusions} contains our conclusions.

\section{Setup of the calculation}
\label{sec:calculation}
In this section we give a brief overview of the computational setup,
and describe the ingredients of both the fixed order
(Section~\ref{sec:FO}) and the resummed (Section~\ref{sec:RES})
calculations.
\subsection{Fixed order}
\label{sec:FO}
For the calculation of the DY process, we consider the off-shell
production of either a pair of charged leptons (mediated by both a $Z$
boson and a virtual photon) or a charged lepton and a neutrino
(mediated by $W^{\pm}$ bosons), in association with partonic jets.
The jet requirement is replaced by a lower cut on the transverse
momentum of the pair, that acts as an infrared regulator of the
fixed-order calculation, hence preventing the radiation from being
entirely unresolved.

The NNLO QCD predictions for neutral and charged DY production have
been obtained in
refs.~\cite{Ridder:2015dxa,Ridder:2016nkl,Gehrmann-DeRidder:2016jns,Gauld:2017tww,Boughezal:2015ded,Boughezal:2016isb
  ,Boughezal:2015dva,Boughezal:2016dtm,Gehrmann-DeRidder:2017mvr}.
Relative to the LO distribution, in which the leptonic system recoils
against a single parton, the NNLO calculation receives contributions
from configurations with two extra partons (RR: double-real
corrections~\cite{Campbell:2002tg,Campbell:2003hd,Hagiwara:1988pp,Berends:1988yn,Falck:1989uz}),
with one extra parton and one extra loop (RV: real-virtual
corrections~\cite{Campbell:2002tg,Campbell:2003hd,Glover:1996eh,Bern:1996ka,Campbell:1997tv,Bern:1997sc})
and with no extra partons but two extra loops (VV: double-virtual
corrections~\cite{Moch:2002hm,Garland:2001tf,Garland:2002ak,Gehrmann:2011ab}).
Each of the three contributions is separately infrared divergent
either in an implicit manner from phase-space regions where the
partonic radiation becomes unresolved (soft and/or collinear), or in a
explicit manner from infrared poles in virtual loop corrections.  Only
the sum of the three contributions is finite.

We perform the calculation using the parton-level generator \nnlojet,
which implements the antenna subtraction
method~\cite{GehrmannDeRidder:2005cm,Daleo:2006xa,Currie:2013vh} to
isolate infrared singularities and to enable their cancellation
between different contributions prior to the numerical phase-space
integration.  The NNLO calculation can be structured as
\begin{align}
\label{eq:NNLOwithansub} 
\sigma^\text{NNLO}_{X+\jet}=&\int_{\Phi_{X+3}}\Big(\rd\sigma^{RR}_\text{NNLO}-\rd\sigma^S_\text{NNLO}\Big)\nonumber\\
+&\int_{\Phi_{X+2}}\Big(\rd\sigma^{RV}_\text{NNLO}-\rd\sigma^T_\text{NNLO}\Big)\nonumber\\
+&\int_{\Phi_{X+1}}\Big(\rd\sigma^{VV}_\text{NNLO}-\rd\sigma^U_\text{NNLO}\Big) .
\end{align} 
The antenna subtraction terms, $\rd\sigma^{S,T,U}_\text{NNLO}$, are
constructed from antenna
functions~\cite{GehrmannDeRidder:2005cm,GehrmannDeRidder:2005aw,Daleo:2009yj,Boughezal:2010mc,Gehrmann:2011wi,GehrmannDeRidder:2012ja}
to cancel infrared singularities between the contributions of
different parton multiplicities. The integrals are performed over the
phase space ${\Phi_{X+1,2,3}}$ corresponding to the production of the
colour singlet in association with one, two or three partons in the
final state.  The integration over the final-state phase space is
fully differential such that any infrared-safe observable $\cal O$ can
be studied through differential distributions as
$\rd\sigma^\text{NNLO}_{X+\jet}/\rd\cal O$.

The matching of the above NNLO prediction to a resummed calculation in
the small $\pt$ limit is computationally very challenging. At small
$\pt$, both the matrix elements and the subtraction terms grow rapidly
in magnitude due to the presence of un-cancelled infrared
singularities. This results in large numerical cancellations between
them that ultimately challenge the stability of the final prediction.
The finite remainder of such cancellations needs to be numerically
stable in order to be consistently combined with a resummed calculation
and extrapolated to the limit $\pt\to 0$. The stability of \nnlojet in
this extreme regime has been tested thoroughly against the expansion
of the N$^3$LL resummations in
refs.~\cite{Chen:2018pzu,Bizon:2018foh}, where it is shown that the
NNLO calculation can be reliably obtained down to very small $\pt$
values.

The residual infrared (logarithmic) divergences that persist in the
$\pt \to 0$ limit are cancelled by combining the fixed-order
computation with a resummed calculation, where the logarithms in the
fixed-order prediction are subtracted and replaced by the sum of
the corresponding enhanced terms to all orders in perturbation theory.
This procedure is discussed in the following Section~\ref{sec:RES}.

\subsection{Resummation and matching}
\label{sec:RES}

The resummation is performed in momentum space by means of the method
formulated in refs.~\cite{Monni:2016ktx,Bizon:2017rah} and implemented
in the computer code \radish. In this approach, the factorisation
properties of the QCD matrix elements in the soft and collinear limits
are exploited to devise a numerical procedure to generate the
radiation at all perturbative orders. This allows for the resummation
of the large logarithmic terms in a fashion that is similar in spirit
to a Monte Carlo generator.  A detailed technical description of the
method can be found in Refs.~\cite{Monni:2016ktx,Bizon:2017rah}, and
the formulae up to N$^3$LL accuracy are collected in
Ref.~\cite{Bizon:2018foh} (Section 3 and Appendix B).

In order to have a reliable prediction across the whole $\pt$
spectrum, the fixed-order and resummed predictions must be
consistently combined through a matching procedure. The matching is
performed in such a way that it reduces to the resummed calculation at
small $\pt$, while reproducing the fixed-order prediction at large
transverse momentum. At a given perturbative order, one can adopt
various schemes that differ from one another by terms beyond the
considered order. In the present analysis we adopt the multiplicative
scheme formulated in refs.~\cite{Caola:2018zye,Bizon:2018foh}, in
which the matching is performed at the level of the cumulative
distribution~\eqref{eq:cumulant-initial} as follows:
\begin{equation}
\label{eq:multiplicative1}
{\rm \Sigma}_{\rm match}^{\rm N^3LL}(\pt) = \frac{{\rm \Sigma}^{\rm N^3LL}(\pt)}{{\rm \Sigma}^{\rm N^3LL}_{\rm asym.} } \left[{\rm \Sigma}^{\rm N^3LL}_{\rm asym.} \frac{{\rm \Sigma}^{\rm N^3LO}(\pt)}{{\rm \Sigma}_{\rm exp.}^{\rm N^3LL}(\pt)}\right]_{\rm N^3LO},
\end{equation}
where ${\rm \Sigma}_{\rm exp.}^{\rm N^3LL}$ denotes the expansion of the resummation
formula ${\rm \Sigma}^{\rm N^3LL}$ to ${\cal O}(\alpha_s^3)$ (N$^3$LO), and
the whole squared bracket in Eq.~\eqref{eq:multiplicative1} is
expanded to N$^3$LO. It should be recalled that ${\cal O}(\alpha_s^3)$  corresponds 
to N$^3$LO in the total (i.e.\ $\pt$-integrated)  cross section and in any 
cumulative distribution, while being 
NNLO in the fixed-order $\pt$-differential cross section. 

In the above equation, ${\rm \Sigma}^{\rm N^3LO}$ is the cumulative fixed-order
distribution at N$^3$LO, i.e.
\begin{equation}
{\rm \Sigma}^{\rm N^3LO}(\pt) = \sigma_{\rm tot}^{\rm N^3LO} - \int_{\pt}^{\infty} \rd \pt' \; \frac{\rd {\rm \Sigma}^{\rm NNLO}(\pt')}{\rd \pt'},
\end{equation}
where $\sigma_{\rm tot}^{\rm N^3LO}$ is the total cross section for
the charged or neutral DY processes at N$^3$LO, and
$\rd {\rm \Sigma}^{\rm NNLO}/\rd \pt'$ denotes the corresponding NNLO
$\pt$-differential distribution obtained with \nnlojet. Both of these
quantities are accurate to ${\cal O}(\alpha_s^3)$.
Since the N$^3$LO inclusive cross section for DY production is
currently unknown, we approximate it with the NNLO cross
section~\cite{Hamberg:1990np,vanNeerven:1991gh,Anastasiou:2003yy,Melnikov:2006di,Melnikov:2006kv,Catani:2010en,Catani:2009sm,Gavin:2010az,Anastasiou:2003ds}
in the following. This approximation impacts only terms at N$^4$LL
order, and is thus beyond the accuracy considered in this study.

Finally, the quantity $ {\rm \Sigma}^{\rm N^3LL}_{\rm asym.}$ is defined as
the asymptotic ($\pt \gg M$) limit of the resummed cross section
\begin{equation}
{\rm \Sigma}^{\rm N^3LL}(\pt)\xrightarrow[\pt \gg M]{} {\rm \Sigma}^{\rm N^3LL}_{\rm asym.} .
\end{equation}
This prescription ensures that, in the $\pt \gg M$ limit,
Eq.~\eqref{eq:multiplicative1} reproduces by construction the
fixed-order result, while in the $\pt \to 0$ limit it reduces to the
resummed prediction. The detailed matching formulae can be found in
Appendix A of ref.~\cite{Bizon:2018foh}.

In the next section, we will also report matched predictions at lower
perturbative orders, NNLL+NLO and NLL+LO, that are obtained from the
following matched cumulative distributions
\begin{align}
\label{eq:multiplicative2}
{\rm \Sigma}_{\rm match}^{\rm NNLL}(\pt) &= \frac{{\rm \Sigma}^{\rm NNLL}(\pt)}{{\rm \Sigma}^{\rm NNLL}_{\rm asym.} } \left[{\rm \Sigma}^{\rm NNLL}_{\rm asym.} \frac{{\rm \Sigma}^{\rm NNLO}(\pt)}{{\rm \Sigma}_{\rm exp.}^{\rm NNLL}(\pt)}\right]_{\rm NNLO}\,,\\
\label{eq:multiplicative3}
{\rm \Sigma}_{\rm match}^{\rm NLL}(\pt) &= \frac{{\rm \Sigma}^{\rm NLL}(\pt)}{{\rm \Sigma}^{\rm NLL}_{\rm asym.} } \left[{\rm \Sigma}^{\rm NLL}_{\rm asym.} \frac{{\rm \Sigma}^{\rm NLO}(\pt)}{{\rm \Sigma}_{\rm exp.}^{\rm NLL}(\pt)}\right]_{\rm NLO}\,.
\end{align}

The above matching schemes guarantee that in the large-$\pt$ limit the
matched cumulative cross sections reproduce, by construction, the
following total cross sections
\begin{align}
{\rm \Sigma}_{\rm match}^{\rm N^3LL}(\pt) &\xrightarrow[\pt \gg M]{}
  \sigma_{\rm tot}^{\rm NNLO}\,,\notag\\
{\rm \Sigma}_{\rm match}^{\rm NNLL}(\pt) &\xrightarrow[\pt \gg M]{}
  \sigma_{\rm tot}^{\rm NNLO}\,,\notag\\
{\rm \Sigma}_{\rm match}^{\rm NLL}(\pt) &\xrightarrow[\pt \gg M]{}
  \sigma_{\rm tot}^{\rm NLO}\,.
\label{eq:normalisation}
\end{align}
We stress once more that the ${\rm \Sigma}_{\rm match}^{\rm N^3LL}$
reproduces the NNLO total cross section at large $\pt$ since the
N$^3$LO result for the inclusive DY process is currently unknown. The
nominal accuracy of the predictions is unaffected by this choice.

The final normalised distributions that will be reported in
Section~\ref{sec:results} are obtained by differentiating
Eqs.~\eqref{eq:multiplicative1},~\eqref{eq:multiplicative2}
and~\eqref{eq:multiplicative3}, and dividing by the respective total
cross sections of the right hand side of Eq.~\eqref{eq:normalisation}.

We recall that the resummed calculation contains a Landau singularity
arising from configurations where the radiation is generated with
transverse momentum scales
$k_\perp\sim M\, \exp\left\{-1/(2\beta_0\alpha_s)\right\}$ (with
$\alpha_s = \alpha_s(M)$ and $\beta_0 = (11\,C_A-2\,n_f)/(12\pi)$).
In the predictions presented in the following, we set the results to
zero when the hardest radiation's transverse momentum reaches the
singularity.  For the leptonic invariant masses studied here, this
procedure acts on radiation emitted at very small transverse momentum
that, due to the vectorial nature of the
observable~\cite{Parisi:1979se,Bizon:2017rah}, gives a very small
contribution to the spectrum. We however stress that for a precise
description of this kinematic regime, a thorough study of the impact
of non-perturbative corrections is necessary.

\section{Results at the LHC}
\label{sec:results}
In this section we report our numerical results for the neutral and
charged DY transverse momentum distributions at N$^3$LL+NNLO.

We consider $pp$ collisions at a centre-of-mass energy of $13~\TeV$,
and we use the NNLO {\tt NNPDF3.1} parton distribution function
set~\cite{Ball:2017nwa} with $\as(M_Z) = 0.118$. The parton densities
are evolved from a low scale $Q_0\sim 1~\GeV$ forwards with {\tt
  LHAPDF}~\cite{Buckley:2014ana}, which correctly implements the heavy
quark thresholds in the PDFs. All convolutions are handled with the
{\tt Hoppet} package~\cite{Salam:2008qg}.
In the results reported below, we use the NNLO DGLAP evolution of the
adopted PDF set for all perturbative orders shown in the
figures. Although the NNLO corrections to the PDF evolution are
formally of order N$^3$LL, we include them also in the NLL and NNLL
predictions in order to guarantee a consistent treatment of the quark
thresholds in the parton densities. We note that this choice will lead
to numerical differences in comparison to other NLL and NNLL results
shown in the literature.

We adopt the $G_\mu$ scheme with the electro-weak parameters taken
from the PDG~\cite{Tanabashi:2018oca}, that is
\begin{align}
& M_Z = 91.1876~\GeV,\quad M_W = 80.379~\GeV,\notag\\
& {\rm \Gamma}_Z = 2.4952~\GeV,\quad {\rm \Gamma}_W = 2.085~\GeV,\notag\\
&\quad G_F=1.1663787\times 10^{-5}~\GeV^{-2}\,.
\end{align}
Moreover, we set the CKM matrix equal to the identity matrix, and we
have verified that this approximation is accurate at the few-permille
level.
For both neutral-current and charged-current DY we apply fiducial
selection cuts that resemble the ones used by ATLAS in previous
analyses~\cite{Aad:2015auj}.

The final state for the neutral DY production is defined by applying
the following set of fiducial selection cuts on the leptonic pair:
\begingroup
\setlength{\jot}{8pt}
\begin{align}
  \label{eq:Z_fiducial}
  &|\vec{p}_{\perp}^{\ell^\pm}|  > 25~\GeV,\quad 
  |\eta^{\ell^\pm}|  < 2.5,\notag\\
  &~ 66~\GeV  < \, \mll \,< 116~\GeV ,
\end{align}
\endgroup 
where $\vec{p}_{\perp}^{\ell^\pm}$ are the transverse momenta of the
two leptons, $\eta^{\ell^\pm}$ are their pseudo-rapidities in the
hadronic centre-of-mass frame, and $\mll$ is the invariant mass of the
di-lepton system.
The central factorisation and renormalisation scales are chosen to be
$\mu_R = \mu_F = \sqrt{\mll^2 + |\vec p_\perp^{Z}|^2 } $ and the
central resummation scale is set to $Q=\mll /2$.

In the case of charged DY production, the fiducial volume is defined as
\begingroup
\setlength{\jot}{8pt}
\begin{align}
  \label{eq:W_fiducial}
  &|\vec{p}_{\perp}^{\ell}|  > 25~\GeV,\quad |\slashed{\vec{E}}_{T}|  > 25~\GeV,\notag\\ 
&~\quad  |\eta^{\ell}|  < 2.5,\quad m_T > 50~\GeV ,
\end{align}
\endgroup 
where $\slashed{\vec{E}}_{T}$ is the missing transverse energy vector and
\begin{equation}
m_T = \sqrt{\left(|\vec{p}_{\perp}^{\ell}| +
    |\slashed{\vec{E}}_{T}|\right)^2 - \left(|\vec{p}_{\perp}^{\ell} +
    \slashed{\vec{E}}_{T}|\right)^2}\,.
\end{equation}
The central factorisation and renormalisation scales are chosen to be
$\mu_R = \mu_F = \sqrt{\mlnu^2 + |\vec p_\perp^{W}|^2 } $ and the
central resummation scale is again set to $Q=\mlnu /2$.

In both processes, we assess the missing higher-order uncertainties by
performing a variation of the renormalisation and factorisation scales
by a factor of two around their respective central values whilst
keeping $1/2 \leq \mu_R/\mu_F \leq 2$.  In addition, for central
factorisation and renormalisation scales, we vary the resummation
scale $Q$ by a factor of two in either direction.  The final
uncertainty is built as the envelope of the resulting nine-scale
variation.

\subsection{Fiducial distributions}
\begin{figure}[htbp]
  \centering
  \includegraphics[trim={0 -0.2cm 0
    0},width=0.9\linewidth]{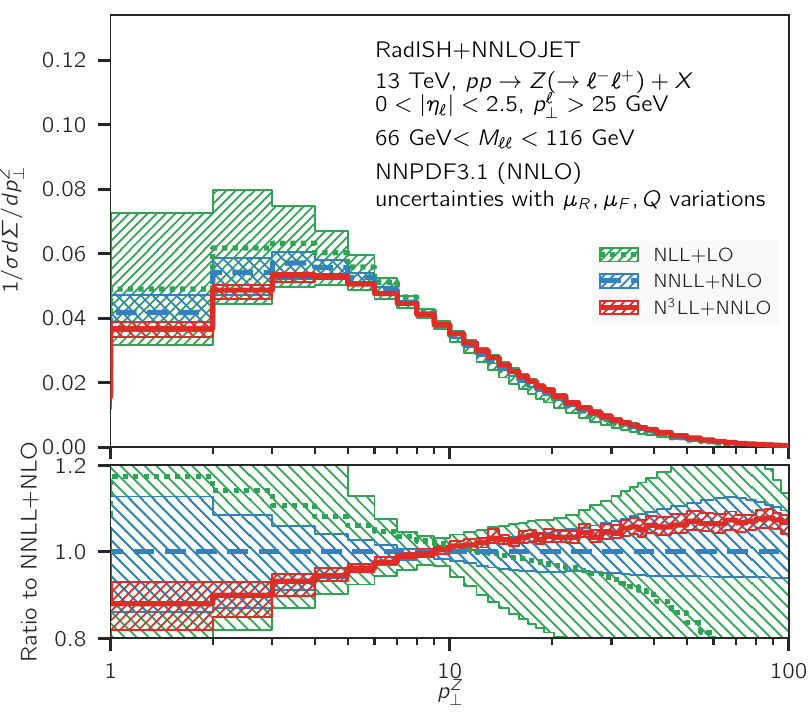} 
  \includegraphics[trim={0 -0.2cm 0
    0},width=0.9\linewidth]{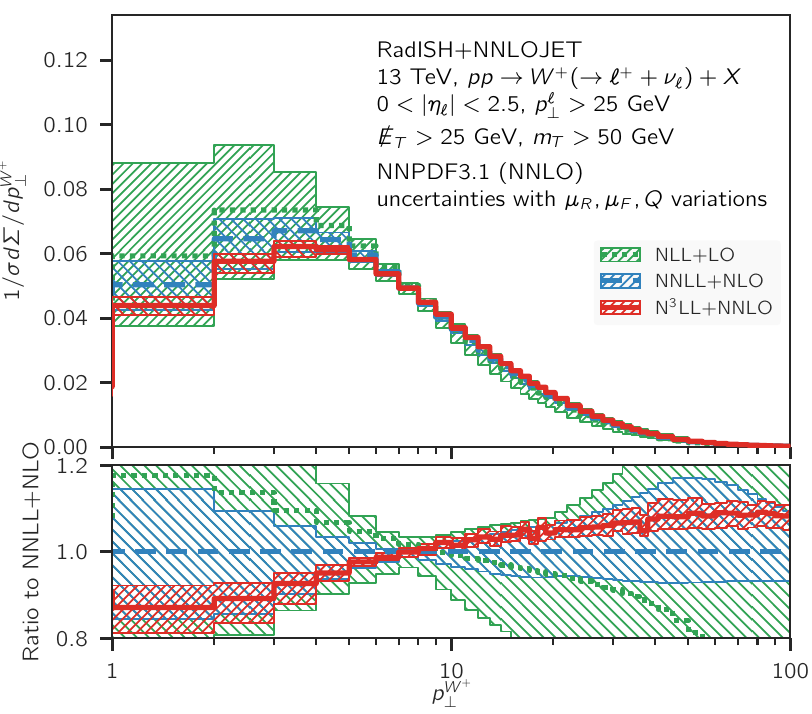} 
  \includegraphics[trim={0 -0.2cm 0
    0},width=0.9\linewidth]{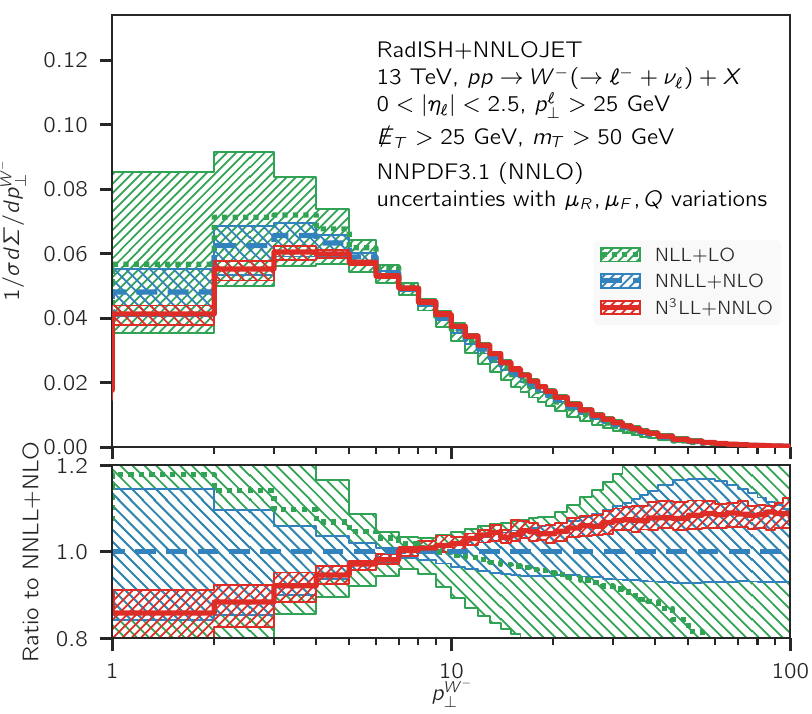} 
  \caption{Comparison of the normalised transverse momentum
    distribution for neutral and charged Drell-Yan pair production at
    NLL+LO (green, dotted), NNLL+NLO (blue, dashed) and N$^3$LL+NNLO (red, solid) at
    $\sqrt{s} = 13~\TeV$ for the fiducial volume defined in the
    text. The lower panel shows the ratio to the NNLL+NLO result.}
  \label{fig:distrib_with_NLL}
\end{figure}

We start by showing, in Figure~\ref{fig:distrib_with_NLL}, the
comparison of the $Z$ and $W^{\pm}$ normalised distributions at NLL+LO
(green), NNLL+NLO (blue), and N$^3$LL+NNLO (red) in the fiducial
volumes defined above. The lower inset of each panel of
Figure~\ref{fig:distrib_with_NLL} shows the ratio of all predictions
to the previous state of the art (NNLL+NLO), with the same colour code
as in the main panels. The difference between each prediction and the
next order is of ${\cal O}(\alpha_s)$, both in the large $\pt$ region
and in the limit $\pt \to 0$ where $\alpha_s L \sim 1$.

In comparison to the NNLL+NLO result, we note that the N$^3$LL+NNLO
corrections lead to important distortions in the shape of the
distributions, making the spectrum harder for $\pt \gtrsim 10~\GeV$,
and softer below this scale.  The perturbative errors are reduced by
more than a factor of two across the whole $\pt$ range, and the
leftover uncertainty is at the $5\%$ level.
In general, we observe a good convergence of the perturbative
description when the order is increased, although in some $\pt$
regions the N$^3$LL+NNLO and the NNLL+NLO bands overlap only
marginally. This feature can be understood by noticing that, as
mentioned in Section~\ref{sec:RES}, both predictions are normalised to
the same NNLO total cross section. Since at large $\pt$ the NNLO
corrections lead to an increase in the spectrum of about $10\%$, by
unitarity of the matching procedure (that preserves the total cross
section) this must be balanced by an analogous decrease in the
spectrum in the region governed by resummation, as indeed observed in
Figure~\ref{fig:distrib_with_NLL}. We stress, nevertheless, that the
two orders are compatible within the quoted uncertainties.

\begin{figure}[htbp]
  \centering
  \includegraphics[trim={0 -0.2cm 0
    0},width=0.9\linewidth]{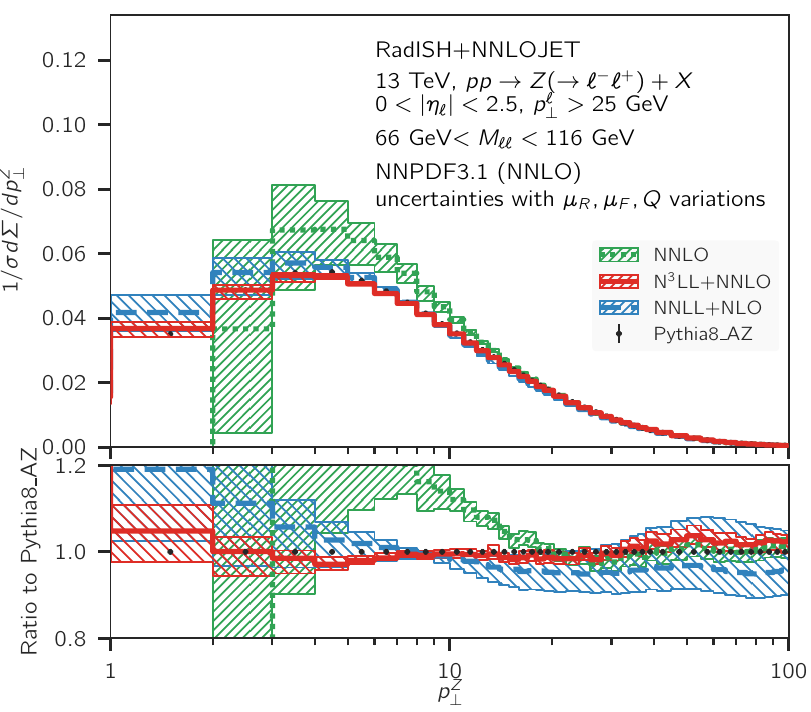} 
  \includegraphics[trim={0 -0.2cm 0
    0},width=0.9\linewidth]{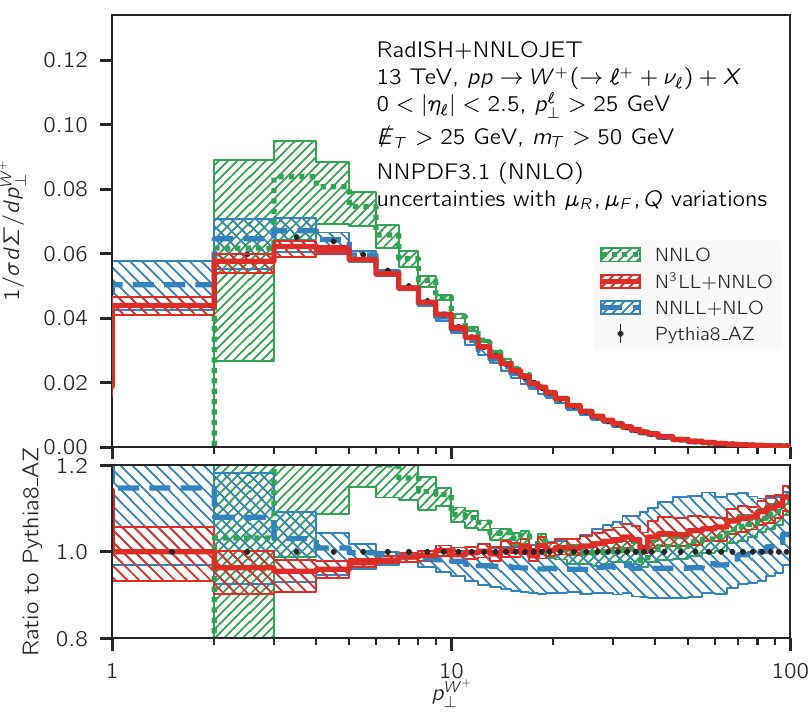} 
  \includegraphics[trim={0 -0.2cm 0 0},width=0.9\linewidth]{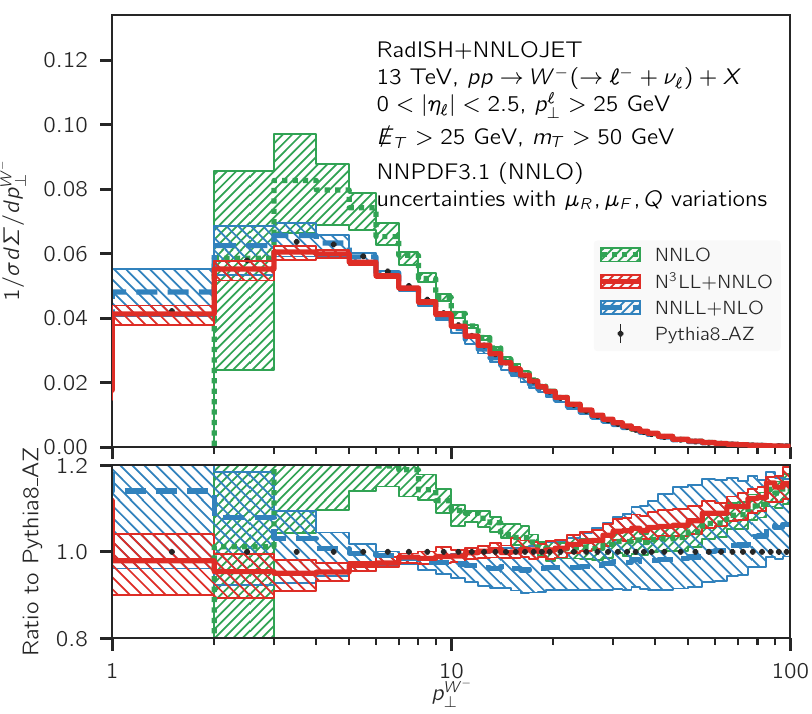} 
  \caption{Comparison of the normalised transverse momentum
    distribution for neutral and charged Drell-Yan pair production at
    NNLO (green, dotted), NNLL+NLO (blue, dashed) and N$^3$LL+NNLO (red, solid) at
    $\sqrt{s} = 13~\TeV$ for the fiducial volume defined in the
    text. For reference, the {\tt Pythia8} prediction in the {\tt AZ}
    tune is also shown, and the lower panel shows the ratio of each
    prediction to the {\tt Pythia8} result.}
  \label{fig:distrib}
\end{figure}

In Figure~\ref{fig:distrib}, we show the comparison among the NNLO
(green), the NNLL+NLO (blue), and N$^3$LL+NNLO (red) predictions,
where the bands are obtained as discussed above. 
Alongside these results, we also show the Monte Carlo predictions
obtained using the \texttt{Pythia8} generator~\cite{Sjostrand:2014zea}
with the {\tt AZ} tune~\cite{Aad:2014xaa}, that has been obtained from
the $Z$-boson $\pt$ distribution at $7~\TeV$. At $7~\TeV$ and $8~\TeV$
the above tune is known to correctly describe the $Z$ transverse
momentum spectrum within a few percent for
$\pt \lesssim 50~\GeV$~\cite{Aad:2014xaa}, and it has been employed in
the extraction of the $W$-boson mass by the ATLAS
collaboration~\cite{Aaboud:2017svj}. Although it is currently unknown
how this tune performs at $13~\TeV$ in comparison to the data, we use
the {\tt Pythia8} prediction for reference in the following plots. In
particular, the lower inset of each panel of Figure~\ref{fig:distrib}
shows the ratio of all predictions to {\tt Pythia8}. We observe a
reasonable agreement between the N$^3$LL+NNLO predictions and {\tt
  Pythia8} below $30~\GeV$, while it deteriorates for larger $\pt$
values. This feature is particularly visible in the case of $W^{\pm}$
production.

\begin{figure}[htbp]
  \centering
  \includegraphics[trim={0 -0.2cm 0
    0},width=0.9\linewidth]{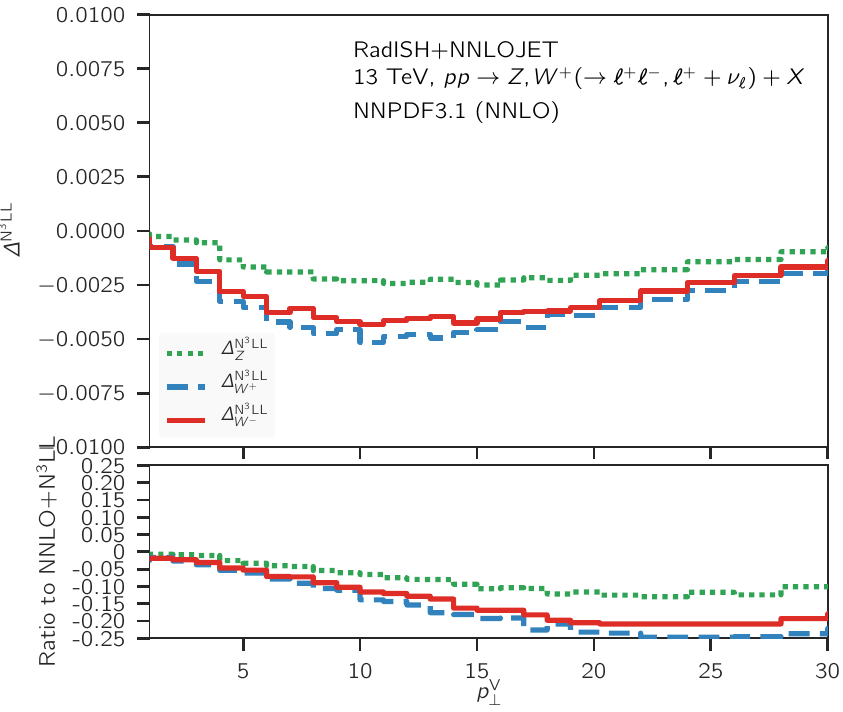} 
  \caption{Difference eq.~\eqref{eq:Delta} between the matched and the resummed predictions for the $Z$ (green, dotted), $W^{+}$ (blue, dashed) and $W^{-}$ (red, solid) normalised distributions. The lower panel shows the ratio of $\Delta^{\rm N^3LL}$  to the N$^3$LL+NNLO matched result.\sloppy}
  \label{fig:nonsingular}
\end{figure}

A comparison of the N$^3$LL+NNLO band to the fixed-order one
shows that the resummation starts making a significant difference for
$\pt \lesssim 20~\GeV$, while above this scale the NNLO provides a
reliable theoretical prediction. 
To further quantify the relative impact of the non-singular contributions in this region, we show  in Fig.~\ref{fig:nonsingular} the difference\sloppy
\begin{equation}\label{eq:Delta}
\Delta^{\rm N^3LL}\equiv( \; \rd \Sigma^{\rm N^3LL+NNLO}/\rd \pt - \; \rd \Sigma^{\rm N^3LL}/\rd \pt )/\sigma_{\rm tot}^{\rm NNLO} 
\end{equation}
between the matched and the resummed predictions for the $Z$ and $W^{\pm}$ normalised distributions.
In the lower panel of the plot we show the relative size of $\Delta^{\rm N^3LL}$ with respect to the matched N$^3$LL+NNLO result, $\Delta^{\rm N^3LL}/( \Sigma^{\rm N^3LL+NNLO}/\rd \pt/ \sigma_{\rm tot}^{\rm NNLO} )$. 
The non-singular contributions are somewhat larger for $W^{\pm}$; the relative size of $\Delta^{\rm N^3LL}$ with respect to the N$^3$LL+NNLO result is  smaller than $5\%$ ($10\%$) for $Z$ ($W^{\pm}$) for $\pt \lesssim 10$~GeV,  and becomes larger than $10\%$ ($20\%$) for $\pt > 20$~GeV. 
\sloppy

\subsection{Ratio of $Z/W$ and $W^-/W^+$ distributions}
Another set of important quantities of interest are the ratios of the
above distributions, which play a central role in recent extractions
of the $W$-boson mass at the LHC~\cite{Aaboud:2017svj}.
When taking ratios of perturbative quantities one has to decide how
to combine the uncertainties in the numerator and denominator to
obtain the final error. 

One option is to try to identify the possible sources of correlation
in the three processes considered here. From the point of view of the
perturbative (massless) QCD description adopted in this study, one
expects the structure of radiative corrections to such reactions to be
nearly identical. This is certainly the case as far as resummation is
concerned, since it is governed by the same anomalous dimensions and
all-order structure in $W$ and $Z$ production. As a consequence, the
resummation scale should be varied in a correlated manner in both
predictions considered in the ratio.
A similar argument can be made regarding the renormalisation scale
$\mu_R$ and the factorisation scale $\mu_F$. 

However, an important difference between $Z$, $W^+$, and $W^-$
production lies in the different combination of partonic channels
probed by each process and, in particular, in the sensitivity to
different heavy quark thresholds in the PDFs at small $\pt$. 
Therefore, it is not clear whether a fully correlated variation of the
factorisation scale $\mu_F$ is physically justified.
A more conservative uncertainty prescription is to vary the scales
$\mu_R$ and $Q$ in numerator and denominator in a fully correlated
way, while varying $\mu_F$ in an uncorrelated manner within the
constraint~\cite{Gehrmann-DeRidder:2017mvr}
\begin{equation}
  \label{eq:constraintA}
\frac{1}{2} \leq \frac{x_{\mu_F}^{\rm num.}}{x_{\mu_F}^{\rm den.}} \leq 2\,,
\end{equation}
where $x_{\mu_F}$ is the ratio of the factorisation scale to its
central value. This corresponds to a total of $17$ scale combinations.

Finally, for comparison we also consider the uncorrelated variation of
$\mu_R$ and $\mu_F$ in the ratio, while imposing
\begin{equation}
  \label{eq:constraintB}
\frac{1}{2} \leq \frac{x_{\mu}^{\rm num.}}{x_{\mu}^{\rm den.}} \leq 2\,,
\end{equation}
where $x_\mu$ is the ratio of the scale $\mu$ to its central value,
with $\mu \equiv \{\mu_R, \mu_F\}$, together with a correlated
variation of the resummation scale $Q$. This recipe amounts to taking
the envelope of the predictions resulting from $33$ different
combinations of scales in the ratio.

\begin{figure}[htbp]
  \centering
  \includegraphics[trim={0 -0.2cm 0
    0},width=0.9\linewidth]{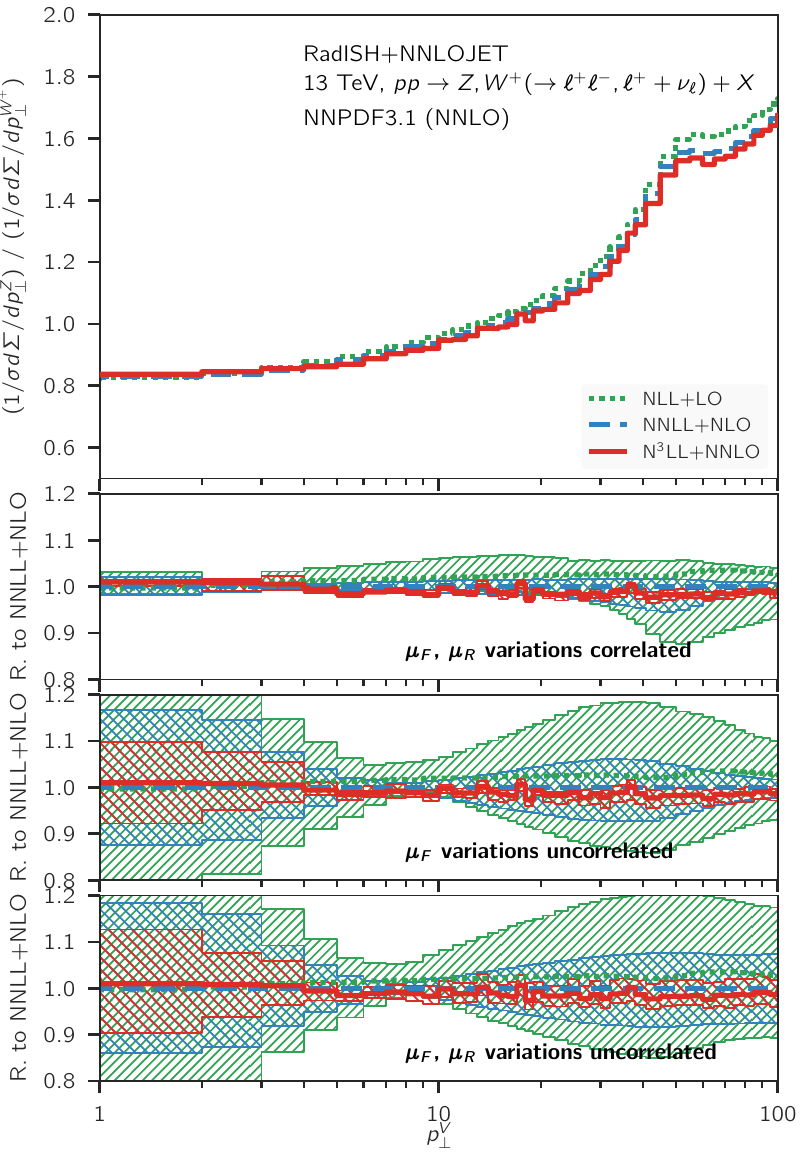} 
  \includegraphics[trim={0 -0.2cm 0
    0},width=0.9\linewidth]{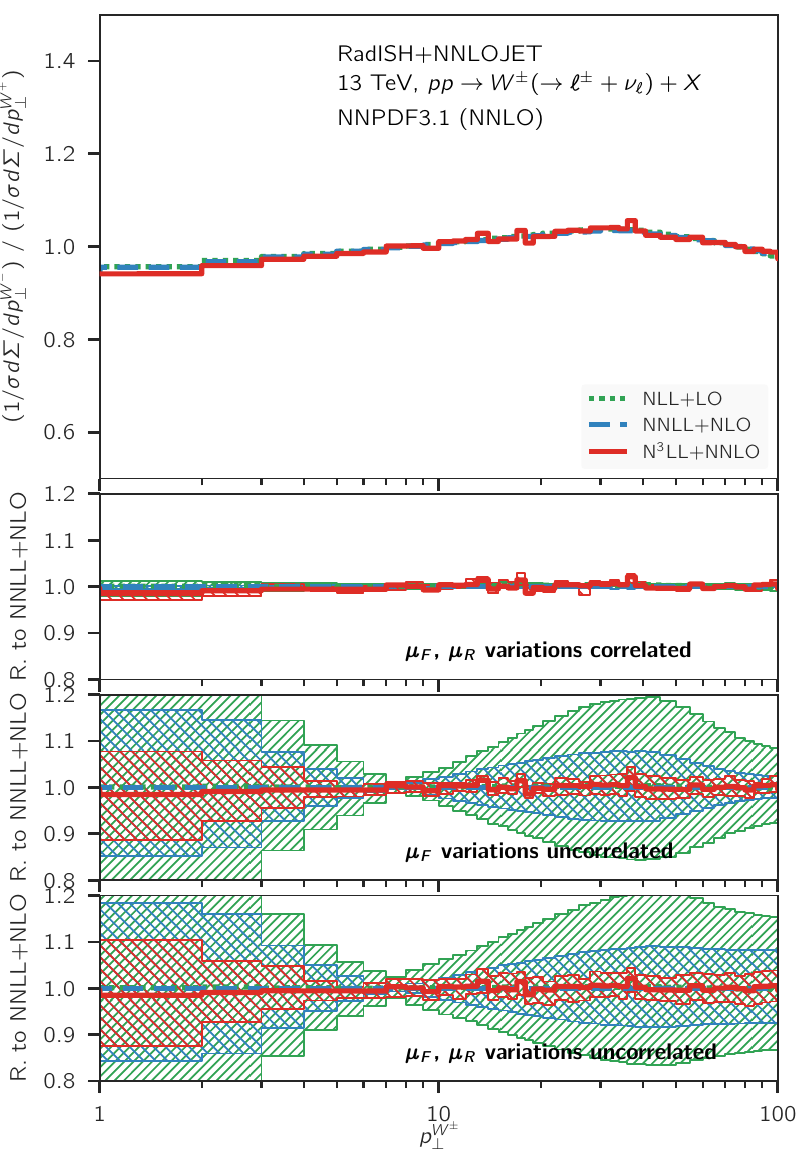} 
  \caption{Ratios of $Z/W^+$ and $W^-/W^+$ normalised differential
    distributions at NLL+LO (green, dotted), NNLL+NLO (blue, dashed) and N$^3$LL+NNLO
    (red, solid) at $\sqrt{s} = 13~\TeV$. The three lower panels show three
    different prescriptions for the theory uncertainty, as described
    in the text.}
  \label{fig:ratios_with_NLL}
\end{figure}

To examine the reliability of the above uncertainty schemes, in
Figure~\ref{fig:ratios_with_NLL} we analyse the convergence of the
perturbative series for the ratios of distributions, by comparing the
results at NLL+LO (green), NNLL+NLO (blue), and N$^3$LL+NNLO
(red). The three lower panels in each plot show the theory
uncertainties obtained according to the three prescriptions outlined
above, respectively, in comparison to the old state-of-the-art
prediction at NNLL+NLO.
In the case of the $Z/W^+$ ratio (shown in the upper plot of
Figure~\ref{fig:ratios_with_NLL}), we observe that the different
perturbative orders are very close to one another. The results are
compatible even within the uncertainty bands obtained with the more
aggressive error estimate, which in some bins is sensitive to
minor statistical fluctuations due to the complexity of the NNLO
calculation.
This feature is strikingly evident in the case of the $W^-/W^+$ ratio
(lower figure), where the excellent convergence of the series seems to
indicate that either a fully correlated scale variation or the more
conservative estimate of Eq.~\eqref{eq:constraintA} is perfectly
justified.\sloppy

\begin{figure}[htbp]
  \centering
  \includegraphics[trim={0 -0.2cm 0
    0},width=0.9\linewidth]{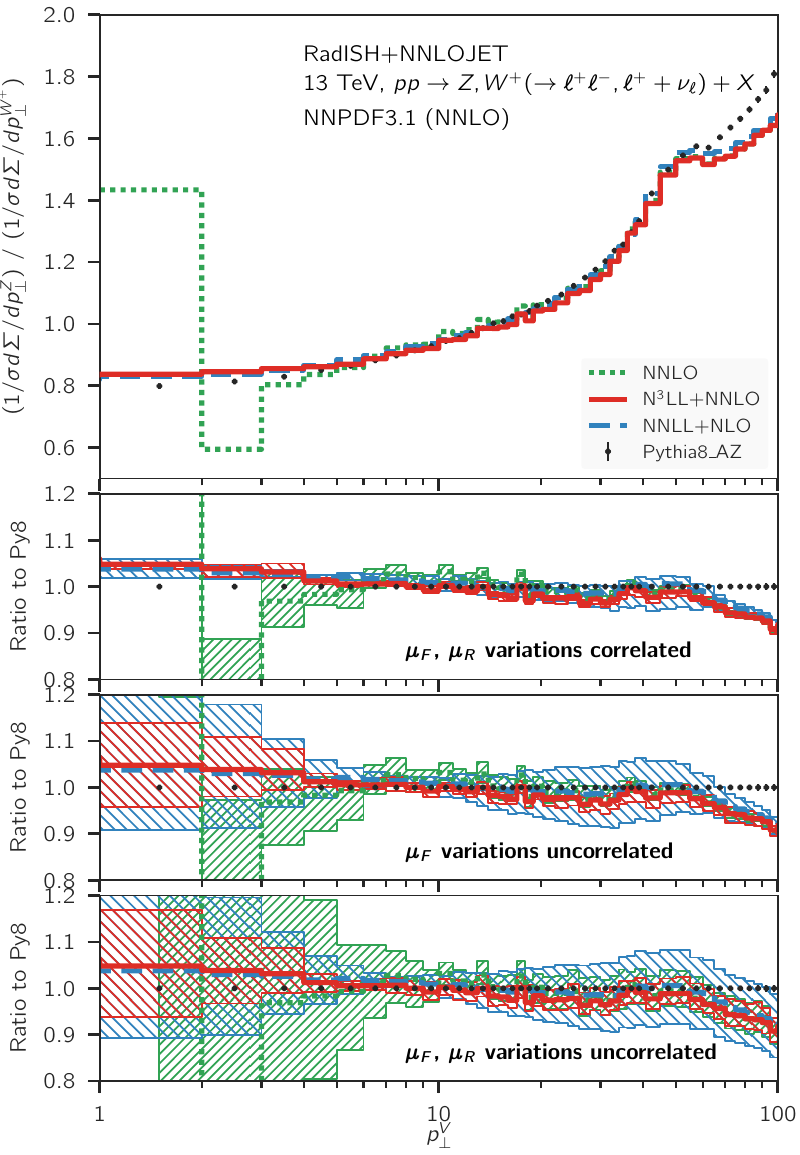} 
  \includegraphics[trim={0 -0.2cm 0
    0},width=0.9\linewidth]{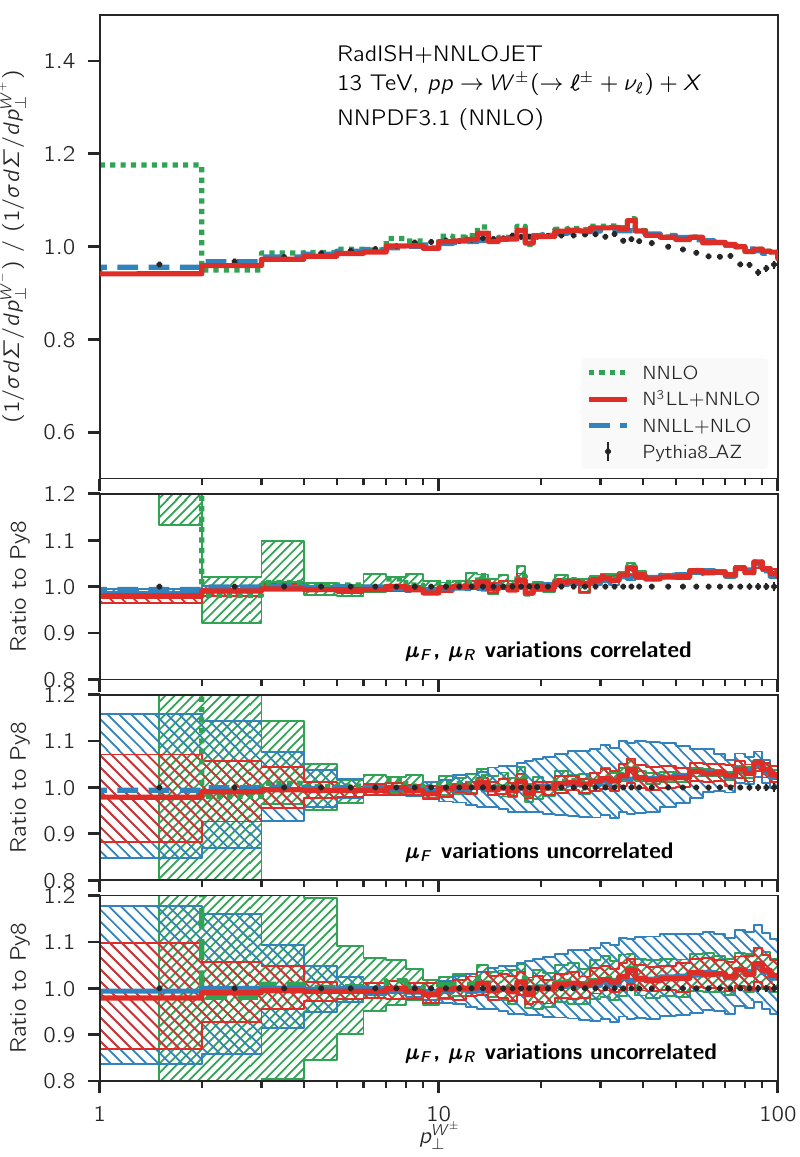} 
  \caption{Ratios of $Z/W^+$ and $W^-/W^+$ normalised differential
    distributions at NNLO (green, dotted), NNLL+NLO (blue, dashed) and N$^3$LL+NNLO
    (red, solid) at $\sqrt{s} = 13~\TeV$. For reference, the {\tt Pythia8}
    prediction in the {\tt AZ} tune is also shown, and the lower panels
    show the ratio of each prediction to the latter.}
  \label{fig:ratios}
\end{figure}

Figure~\ref{fig:ratios} shows the comparison of the same two ratios
($Z/W^+$ and $W^-/W^+$) to the NNLO result (green), and to {\tt
  Pythia8}.
We observe that in both cases the N$^3$LL+NNLO calculation leads to an
important reduction of the theory uncertainty. In particular, even
with the most conservative estimate of the theory error, our best
prediction leads to errors of the order of $5\%$, with the exception
of the first bin where the perturbative uncertainty is at the $10\%$
level. The kink around $\pt \sim 50-60~\GeV$ in the $Z/W^+$ ratio
(upper plot in Figure~\ref{fig:ratios}) is due to the different
fiducial selection cuts in the two processes. A change in the shape of
the distributions around this scale is indeed visible in
Figure~\ref{fig:distrib}, at slightly different $\pt$ values for $Z$
and $W^+$ production, respectively, that is reflected in the structure
observed in Figure~\ref{fig:ratios}.
We find a good agreement between our best predictions at N$^3$LL+NNLO
and the {\tt Pythia8} Monte Carlo in the small $\pt$ region of the
ratios. However, the two predictions are not compatible within the
quoted theory uncertainties if the scales are varied in a fully
correlated manner. On the other hand, for $\pt\gtrsim 40~\GeV$, the
{\tt Pythia8} result disagrees with the matched calculation.
This behavior is not unexpected, since the nominal perturbative
accuracy of {\tt Pythia8} is well below any of the matched
calculations, and the {\tt AZ} tune is optimised to describe the $Z$
spectrum in the region $\pt \leq 50~\GeV$ at $7~\TeV$.

\section{Conclusions}
\label{sec:conclusions}

In this work, we computed the transverse momentum distributions of
electroweak gauge bosons at the LHC to N$^3$LL+NNLO accuracy in QCD.
This calculation opens up a new level of theoretical precision in the
description of these observables. The new state-of-the-art prediction
is obtained by combining the NNLO results from the \nnlojet program
with the N$^3$LL resummation performed with \radish.
Our phenomenological study adopts fiducial selection cuts similar to
the setup adopted by ATLAS in previous studies.
The numerical results we presented are made available in electronic format
as additional material alongside this manuscript.

We find that, in comparison to the fixed-order prediction, the
resummation effects become important for $\pt \lesssim 20~\GeV$.
The effect of the N$^3$LL+NNLO corrections with respect to the
previous NNLL+NLO prediction is as large as $\sim 10\%$, and leads to
significant shape distortions as well as to a substantial reduction in
the perturbative uncertainty due to missing higher-order
corrections. In particular, the distributions considered in this
article are predicted with a residual uncertainty below the $5\%$
level across most of the $\pt$ spectrum.
We also compared the results to the predictions obtained from the {\tt
  Pythia8} Monte Carlo with the {\tt AZ} tune, that has been
determined using the ATLAS experimental data for the $Z$ boson
transverse momentum at $7~\TeV$~\cite{Aad:2014xaa}.

Finally, we examined the ratios of the $Z$ to $W^+$, and $W^-$ to
$W^+$ distributions, which play an important role in the $W$ mass
extraction at the LHC. We consider different prescriptions for the
estimate of perturbative uncertainties that rely on different degrees
of correlation between the scales in the numerator and in the
denominator. We find a remarkable convergence of the predictions for
the ratios at different perturbative orders. This fact strongly
indicates that the class of processes considered in this study feature
very similar perturbative corrections suggesting that the perturbative
sources of uncertainty are correlated to a large extent.

There are, however, additional sources of perturbative corrections to
$W^{\pm}$ and $Z$ production that we ignored in our study. In
particular, at the level of the residual theoretical errors obtained
in our predictions, PDF theory uncertainties~\cite{Harland-Lang:2018bxd,AbdulKhalek:2019bux}, QED
corrections~\cite{deFlorian:2018wcj,Cieri:2018sfk}, as well as a
careful study of the impact of mass
effects~\cite{Aivazis:1993pi,Thorne:1997ga,Collins:1998rz,Berge:2005rv,Forte:2010ta,Pietrulewicz:2017gxc,CarloniCalame:2016ouw,Krauss:2017wmx,Figueroa:2018chn,Forte:2018ovl,Bagnaschi:2018dnh}
become necessary. The correlation pattern between the uncertainties
due to such effects may well be different from what we have observed
in this paper, and a dedicated study must be performed in order to
reliably combine these effects with the N$^3$LL+NNLO predictions
presented here.

\section*{Acknowledgements}
We are very grateful to Jan Kretzschmar for clarifying discussions and
for kindly providing us with the {\tt Pythia8} setup used by ATLAS and
with template predictions for the $W$ and $Z$ differential
distributions. We would like to thank Paolo Torrielli for fruitful
discussions on the topics presented here, and Stefano Camarda for
useful correspondence on the impact of heavy flavours. We also thank
Fabrizio Caola, Lucian Harland-Lang, Jonas Lindert, Gavin Salam, and
Marek Sch\"onherr for constructive comments on the uncertainty
prescriptions. The work of PM is supported by the Marie Sk\l{}odowska
Curie Individual Fellowship contract number 702610 Resummation4PS. LR
is supported by the ERC Starting Grant REINVENT (714788) and
acknowledges the CERN Theoretical Physics Department for hospitality
and support during part of this work. This work has been supported in
part by the Swiss National Science Foundation (SNF) under grant
numbers 200021-172478 and 200020-175595, and by the Research Executive
Agency (REA) of the European Union through the ERC Consolidator Grant
HICCUP (614577) and the ERC Advanced Grant MC@NNLO (340983).

\bibliographystyle{spphys}
\bibliography{biblio}

\begin{thebibliography}{100}
\providecommand{\url}[1]{{#1}}
\providecommand{\urlprefix}{URL }
\expandafter\ifx\csname urlstyle\endcsname\relax
  \providecommand{\doi}[1]{DOI \discretionary{}{}{}#1}\else
  \providecommand{\doi}{DOI \discretionary{}{}{}\begingroup
  \urlstyle{rm}\Url}\fi

\bibitem{Chatrchyan:2011wt}
S.~Chatrchyan, et~al., Phys. Rev. \textbf{D85}, 032002 (2012), 1110.4973

\bibitem{Aad:2012wfa}
G.~Aad, et~al., Phys. Lett. \textbf{B720}, 32 (2013), 1211.6899

\bibitem{Aad:2014xaa}
G.~Aad, et~al., JHEP \textbf{09}, 145 (2014), 1406.3660

\bibitem{Aad:2015auj}
G.~Aad, et~al., Eur. Phys. J. \textbf{C76}(5), 291 (2016), 1512.02192

\bibitem{Aaij:2015gna}
R.~Aaij, et~al., JHEP \textbf{08}, 039 (2015), 1505.07024

\bibitem{Khachatryan:2015oaa}
V.~Khachatryan, et~al., Phys. Lett. \textbf{B749}, 187 (2015), 1504.03511

\bibitem{Khachatryan:2015paa}
V.~Khachatryan, et~al., Phys. Lett. \textbf{B750}, 154 (2015), 1504.03512

\bibitem{Aaij:2015zlq}
R.~Aaij, et~al., JHEP \textbf{01}, 155 (2016), 1511.08039

\bibitem{Aad:2016izn}
G.~Aad, et~al., JHEP \textbf{08}, 159 (2016), 1606.00689

\bibitem{Khachatryan:2016nbe}
V.~Khachatryan, et~al., JHEP \textbf{02}, 096 (2017), 1606.05864

\bibitem{Aaij:2016mgv}
R.~Aaij, et~al., JHEP \textbf{09}, 136 (2016), 1607.06495

\bibitem{Sirunyan:2017igm}
A.M. Sirunyan, et~al., JHEP \textbf{03}, 172 (2018), 1710.07955

\bibitem{Aaboud:2017svj}
M.~Aaboud, et~al., Eur. Phys. J. \textbf{C78}(2), 110 (2018), 1701.07240

\bibitem{Aaboud:2017ffb}
M.~Aaboud, et~al., JHEP \textbf{12}, 059 (2017), 1710.05167

\bibitem{Ball:2017nwa}
R.D. Ball, et~al., Eur. Phys. J. \textbf{C77}(10), 663 (2017), 1706.00428

\bibitem{Boughezal:2017nla}
R.~Boughezal, A.~Guffanti, F.~Petriello, M.~Ubiali, JHEP \textbf{07}, 130
  (2017), 1705.00343

\bibitem{Bacchetta:2017gcc}
A.~Bacchetta, F.~Delcarro, C.~Pisano, M.~Radici, A.~Signori, JHEP \textbf{06},
  081 (2017), 1703.10157

\bibitem{Bertone:2019nxa}
V.~Bertone, I.~Scimemi, A.~Vladimirov,   (2019), 1902.08474

\bibitem{Hamberg:1990np}
R.~Hamberg, W.L. van Neerven, T.~Matsuura, Nucl. Phys. \textbf{B359}, 343
  (1991).
\newblock [Erratum: Nucl. Phys.B644,403(2002)]

\bibitem{vanNeerven:1991gh}
W.L. van Neerven, E.B. Zijlstra, Nucl. Phys. \textbf{B382}, 11 (1992).
\newblock [Erratum: Nucl. Phys.B680,513(2004)]

\bibitem{Anastasiou:2003yy}
C.~Anastasiou, L.J. Dixon, K.~Melnikov, F.~Petriello, Phys. Rev. Lett.
  \textbf{91}, 182002 (2003), hep-ph/0306192

\bibitem{Melnikov:2006di}
K.~Melnikov, F.~Petriello, Phys. Rev. Lett. \textbf{96}, 231803 (2006),
  hep-ph/0603182

\bibitem{Melnikov:2006kv}
K.~Melnikov, F.~Petriello, Phys. Rev. \textbf{D74}, 114017 (2006),
  hep-ph/0609070

\bibitem{Catani:2010en}
S.~Catani, G.~Ferrera, M.~Grazzini, JHEP \textbf{05}, 006 (2010), 1002.3115

\bibitem{Catani:2009sm}
S.~Catani, L.~Cieri, G.~Ferrera, D.~de~Florian, M.~Grazzini, Phys. Rev. Lett.
  \textbf{103}, 082001 (2009), 0903.2120

\bibitem{Gavin:2010az}
R.~Gavin, Y.~Li, F.~Petriello, S.~Quackenbush, Comput. Phys. Commun.
  \textbf{182}, 2388 (2011), 1011.3540

\bibitem{Anastasiou:2003ds}
C.~Anastasiou, L.J. Dixon, K.~Melnikov, F.~Petriello, Phys. Rev. \textbf{D69},
  094008 (2004), hep-ph/0312266

\bibitem{Ridder:2015dxa}
A.~Gehrmann-De~Ridder, T.~Gehrmann, E.W.N. Glover, A.~Huss, T.A. Morgan, Phys.
  Rev. Lett. \textbf{117}(2), 022001 (2016), 1507.02850

\bibitem{Ridder:2016nkl}
A.~Gehrmann-De~Ridder, T.~Gehrmann, E.W.N. Glover, A.~Huss, T.A. Morgan, JHEP
  \textbf{07}, 133 (2016), 1605.04295

\bibitem{Gehrmann-DeRidder:2016jns}
A.~Gehrmann-De~Ridder, T.~Gehrmann, E.W.N. Glover, A.~Huss, T.A. Morgan, JHEP
  \textbf{11}, 094 (2016), 1610.01843

\bibitem{Gauld:2017tww}
R.~Gauld, A.~Gehrmann-De~Ridder, T.~Gehrmann, E.W.N. Glover, A.~Huss, JHEP
  \textbf{11}, 003 (2017), 1708.00008

\bibitem{Boughezal:2015ded}
R.~Boughezal, J.M. Campbell, R.K. Ellis, C.~Focke, W.T. Giele, X.~Liu,
  F.~Petriello, Phys. Rev. Lett. \textbf{116}(15), 152001 (2016), 1512.01291

\bibitem{Boughezal:2016isb}
R.~Boughezal, X.~Liu, F.~Petriello, Phys. Rev. \textbf{D94}(7), 074015 (2016),
  1602.08140

\bibitem{Boughezal:2015dva}
R.~Boughezal, C.~Focke, X.~Liu, F.~Petriello, Phys. Rev. Lett. \textbf{115}(6),
  062002 (2015), 1504.02131

\bibitem{Boughezal:2016dtm}
R.~Boughezal, X.~Liu, F.~Petriello, Phys. Rev. \textbf{D94}(11), 113009 (2016),
  1602.06965

\bibitem{Gehrmann-DeRidder:2017mvr}
A.~Gehrmann-De~Ridder, T.~Gehrmann, E.W.N. Glover, A.~Huss, D.M. Walker, Phys.
  Rev. Lett. \textbf{120}(12), 122001 (2018), 1712.07543

\bibitem{Kuhn:2005az}
J.H. K{\"u}hn, A.~Kulesza, S.~Pozzorini, M.~Schulze, Nucl. Phys. \textbf{B727},
  368 (2005), hep-ph/0507178

\bibitem{Kuhn:2007qc}
J.H. K{\"u}hn, A.~Kulesza, S.~Pozzorini, M.~Schulze, Phys. Lett. \textbf{B651},
  160 (2007), hep-ph/0703283

\bibitem{Denner:2009gj}
A.~Denner, S.~Dittmaier, T.~Kasprzik, A.~M{\"u}ck, JHEP \textbf{08}, 075
  (2009), 0906.1656

\bibitem{Denner:2011vu}
A.~Denner, S.~Dittmaier, T.~Kasprzik, A.~M{\"u}ck, JHEP \textbf{06}, 069
  (2011), 1103.0914

\bibitem{Parisi:1979se}
G.~Parisi, R.~Petronzio, Nucl. Phys. \textbf{B154}, 427 (1979)

\bibitem{Collins:1984kg}
J.C. Collins, D.E. Soper, G.F. Sterman, Nucl. Phys. \textbf{B250}, 199 (1985)

\bibitem{Balazs:1997xd}
C.~Balazs, C.P. Yuan, Phys. Rev. \textbf{D56}, 5558 (1997), hep-ph/9704258

\bibitem{Landry:2002ix}
F.~Landry, R.~Brock, P.M. Nadolsky, C.P. Yuan, Phys. Rev. \textbf{D67}, 073016
  (2003), hep-ph/0212159

\bibitem{Becher:2010tm}
T.~Becher, M.~Neubert, Eur. Phys. J. \textbf{C71}, 1665 (2011), 1007.4005

\bibitem{Bozzi:2010xn}
G.~Bozzi, S.~Catani, G.~Ferrera, D.~de~Florian, M.~Grazzini, Phys. Lett.
  \textbf{B696}, 207 (2011), 1007.2351

\bibitem{Becher:2011xn}
T.~Becher, M.~Neubert, D.~Wilhelm, JHEP \textbf{02}, 124 (2012), 1109.6027

\bibitem{GarciaEchevarria:2011rb}
M.G. Echevarria, A.~Idilbi, I.~Scimemi, JHEP \textbf{07}, 002 (2012), 1111.4996

\bibitem{Monni:2016ktx}
P.F. Monni, E.~Re, P.~Torrielli, Phys. Rev. Lett. \textbf{116}(24), 242001
  (2016), 1604.02191

\bibitem{Ebert:2016gcn}
M.A. Ebert, F.J. Tackmann, JHEP \textbf{02}, 110 (2017), 1611.08610

\bibitem{Bizon:2017rah}
W.~Bizon, P.F. Monni, E.~Re, L.~Rottoli, P.~Torrielli, JHEP \textbf{02}, 108
  (2018), 1705.09127

\bibitem{Chen:2018pzu}
X.~Chen, T.~Gehrmann, E.W.N. Glover, A.~Huss, Y.~Li, D.~Neill, M.~Schulze, I.W.
  Stewart, H.X. Zhu, Phys. Lett. \textbf{B788}, 425 (2019), 1805.00736

\bibitem{Bizon:2018foh}
W.~Bizoń, X.~Chen, A.~Gehrmann-De~Ridder, T.~Gehrmann, N.~Glover, A.~Huss,
  P.F. Monni, E.~Re, L.~Rottoli, P.~Torrielli, JHEP \textbf{12}, 132 (2018),
  1805.05916

\bibitem{Becher:2019bnm}
T.~Becher, M.~Hager,   (2019), 1904.08325

\bibitem{Catani:2012qa}
S.~Catani, L.~Cieri, D.~de~Florian, G.~Ferrera, M.~Grazzini, Eur. Phys. J.
  \textbf{C72}, 2195 (2012), 1209.0158

\bibitem{Gehrmann:2014yya}
T.~Gehrmann, T.~Lübbert, L.L. Yang, JHEP \textbf{06}, 155 (2014), 1403.6451

\bibitem{Echevarria:2016scs}
M.G. Echevarria, I.~Scimemi, A.~Vladimirov, JHEP \textbf{09}, 004 (2016),
  1604.07869

\bibitem{Li:2016ctv}
Y.~Li, H.X. Zhu, Phys. Rev. Lett. \textbf{118}(2), 022004 (2017), 1604.01404

\bibitem{Vladimirov:2016dll}
A.A. Vladimirov, Phys. Rev. Lett. \textbf{118}(6), 062001 (2017), 1610.05791

\bibitem{Moch:2018wjh}
S.~Moch, B.~Ruijl, T.~Ueda, J.A.M. Vermaseren, A.~Vogt, Phys. Lett.
  \textbf{B782}, 627 (2018), 1805.09638

\bibitem{Lee:2019zop}
R.N. Lee, A.V. Smirnov, V.A. Smirnov, M.~Steinhauser, JHEP \textbf{02}, 172
  (2019), 1901.02898

\bibitem{Campbell:2002tg}
J.M. Campbell, R.K. Ellis, Phys. Rev. \textbf{D65}, 113007 (2002),
  hep-ph/0202176

\bibitem{Campbell:2003hd}
J.M. Campbell, R.K. Ellis, D.L. Rainwater, Phys. Rev. \textbf{D68}, 094021
  (2003), hep-ph/0308195

\bibitem{Hagiwara:1988pp}
K.~Hagiwara, D.~Zeppenfeld, Nucl. Phys. \textbf{B313}, 560 (1989)

\bibitem{Berends:1988yn}
F.A. Berends, W.T. Giele, H.~Kuijf, Nucl. Phys. \textbf{B321}, 39 (1989)

\bibitem{Falck:1989uz}
N.K. Falck, D.~Graudenz, G.~Kramer, Nucl. Phys. \textbf{B328}, 317 (1989)

\bibitem{Glover:1996eh}
E.W.N. Glover, D.J. Miller, Phys. Lett. \textbf{B396}, 257 (1997),
  hep-ph/9609474

\bibitem{Bern:1996ka}
Z.~Bern, L.J. Dixon, D.A. Kosower, S.~Weinzierl, Nucl. Phys. \textbf{B489}, 3
  (1997), hep-ph/9610370

\bibitem{Campbell:1997tv}
J.M. Campbell, E.W.N. Glover, D.J. Miller, Phys. Lett. \textbf{B409}, 503
  (1997), hep-ph/9706297

\bibitem{Bern:1997sc}
Z.~Bern, L.J. Dixon, D.A. Kosower, Nucl. Phys. \textbf{B513}, 3 (1998),
  hep-ph/9708239

\bibitem{Moch:2002hm}
S.~Moch, P.~Uwer, S.~Weinzierl, Phys. Rev. \textbf{D66}, 114001 (2002),
  hep-ph/0207043

\bibitem{Garland:2001tf}
L.W. Garland, T.~Gehrmann, E.W.N. Glover, A.~Koukoutsakis, E.~Remiddi, Nucl.
  Phys. \textbf{B627}, 107 (2002), hep-ph/0112081

\bibitem{Garland:2002ak}
L.W. Garland, T.~Gehrmann, E.W.N. Glover, A.~Koukoutsakis, E.~Remiddi, Nucl.
  Phys. \textbf{B642}, 227 (2002), hep-ph/0206067

\bibitem{Gehrmann:2011ab}
T.~Gehrmann, L.~Tancredi, JHEP \textbf{02}, 004 (2012), 1112.1531

\bibitem{GehrmannDeRidder:2005cm}
A.~Gehrmann-De~Ridder, T.~Gehrmann, E.W.N. Glover, JHEP \textbf{0509}, 056
  (2005), hep-ph/0505111

\bibitem{Daleo:2006xa}
A.~Daleo, T.~Gehrmann, D.~Maitre, JHEP \textbf{04}, 016 (2007), hep-ph/0612257

\bibitem{Currie:2013vh}
J.~Currie, E.W.N. Glover, S.~Wells, JHEP \textbf{04}, 066 (2013), 1301.4693

\bibitem{GehrmannDeRidder:2005aw}
A.~Gehrmann-De~Ridder, T.~Gehrmann, E.W.N. Glover, Phys. Lett. \textbf{B612},
  49 (2005), hep-ph/0502110

\bibitem{Daleo:2009yj}
A.~Daleo, A.~Gehrmann-De~Ridder, T.~Gehrmann, G.~Luisoni, JHEP \textbf{1001},
  118 (2010), 0912.0374

\bibitem{Boughezal:2010mc}
R.~Boughezal, A.~Gehrmann-De~Ridder, M.~Ritzmann, JHEP \textbf{02}, 098 (2011),
  1011.6631

\bibitem{Gehrmann:2011wi}
T.~Gehrmann, P.F. Monni, JHEP \textbf{12}, 049 (2011), 1107.4037

\bibitem{GehrmannDeRidder:2012ja}
A.~Gehrmann-De~Ridder, T.~Gehrmann, M.~Ritzmann, JHEP \textbf{10}, 047 (2012),
  1207.5779

\bibitem{Caola:2018zye}
F.~Caola, J.M. Lindert, K.~Melnikov, P.F. Monni, L.~Tancredi, C.~Wever, JHEP
  \textbf{09}, 035 (2018), 1804.07632

\bibitem{Buckley:2014ana}
A.~Buckley, J.~Ferrando, S.~Lloyd, K.~Nordström, B.~Page, M.~Rüfenacht,
  M.~Schönherr, G.~Watt, Eur. Phys. J. \textbf{C75}, 132 (2015), 1412.7420

\bibitem{Salam:2008qg}
G.P. Salam, J.~Rojo, Comput. Phys. Commun. \textbf{180}, 120 (2009), 0804.3755

\bibitem{Tanabashi:2018oca}
M.~Tanabashi, et~al., Phys. Rev. \textbf{D98}(3), 030001 (2018)

\bibitem{Sjostrand:2014zea}
T.~Sjöstrand, S.~Ask, J.R. Christiansen, R.~Corke, N.~Desai, P.~Ilten,
  S.~Mrenna, S.~Prestel, C.O. Rasmussen, P.Z. Skands, Comput. Phys. Commun.
  \textbf{191}, 159 (2015), 1410.3012

\bibitem{Harland-Lang:2018bxd}
L.A. Harland-Lang, R.S. Thorne, Eur. Phys. J. \textbf{C79}(3), 225 (2019),
  1811.08434

\bibitem{AbdulKhalek:2019bux}
R.~Abdul~Khalek, et~al., Eur. Phys. J. \textbf{C79}, 838 (2019), 1905.04311

\bibitem{deFlorian:2018wcj}
D.~de~Florian, M.~Der, I.~Fabre, Phys. Rev. \textbf{D98}(9), 094008 (2018),
  1805.12214

\bibitem{Cieri:2018sfk}
L.~Cieri, G.~Ferrera, G.F.R. Sborlini, JHEP \textbf{08}, 165 (2018), 1805.11948

\bibitem{Aivazis:1993pi}
M.A.G. Aivazis, J.C. Collins, F.I. Olness, W.K. Tung, Phys. Rev. \textbf{D50},
  3102 (1994), hep-ph/9312319

\bibitem{Thorne:1997ga}
R.S. Thorne, R.G. Roberts, Phys. Rev. \textbf{D57}, 6871 (1998), hep-ph/9709442

\bibitem{Collins:1998rz}
J.C. Collins, Phys. Rev. \textbf{D58}, 094002 (1998), hep-ph/9806259

\bibitem{Berge:2005rv}
S.~Berge, P.M. Nadolsky, F.I. Olness, Phys. Rev. \textbf{D73}, 013002 (2006),
  hep-ph/0509023

\bibitem{Forte:2010ta}
S.~Forte, E.~Laenen, P.~Nason, J.~Rojo, Nucl. Phys. \textbf{B834}, 116 (2010),
  1001.2312

\bibitem{Pietrulewicz:2017gxc}
P.~Pietrulewicz, D.~Samitz, A.~Spiering, F.J. Tackmann, JHEP \textbf{08}, 114
  (2017), 1703.09702

\bibitem{CarloniCalame:2016ouw}
C.M. Carloni~Calame, M.~Chiesa, H.~Martinez, G.~Montagna, O.~Nicrosini,
  F.~Piccinini, A.~Vicini, Phys. Rev. \textbf{D96}(9), 093005 (2017),
  1612.02841

\bibitem{Krauss:2017wmx}
F.~Krauss, D.~Napoletano, Phys. Rev. \textbf{D98}(9), 096002 (2018), 1712.06832

\bibitem{Figueroa:2018chn}
D.~Figueroa, S.~Honeywell, S.~Quackenbush, L.~Reina, C.~Reuschle, D.~Wackeroth,
  Phys. Rev. \textbf{D98}(9), 093002 (2018), 1805.01353

\bibitem{Forte:2018ovl}
S.~Forte, D.~Napoletano, M.~Ubiali, Eur. Phys. J. \textbf{C78}(11), 932 (2018),
  1803.10248

\bibitem{Bagnaschi:2018dnh}
E.~Bagnaschi, F.~Maltoni, A.~Vicini, M.~Zaro, JHEP \textbf{07}, 101 (2018),
  1803.04336

\end{thebibliography}

\end{document}